\documentclass[final,authoryear,3p,times,twocolumn]{elsarticle}

\usepackage[latin2]{inputenc}
\usepackage{graphicx}

\usepackage{amssymb}
\usepackage{amsthm}
\usepackage{ulem}

\journal{Applied Radiation and Isotopes}

\begin{document}

\begin{frontmatter}



\title{Extension of the energy range of the experimental activation cross-sections data of longer-lived products of proton induced nuclear reactions on dysprosium up to 65 MeV}


\author[1]{F. T\'ark\'anyi}
\author[1]{F. Ditr\'oi\corref{*}}
\author[1]{S. Tak\'acs}
\author[2]{A. Hermanne}
\author[3]{A.V. Ignatyuk}

\cortext[*]{Corresponding author: ditroi@atomki.hu}

\address[1]{Institute for Nuclear Research, Hungarian Academy of Sciences (ATOMKI),  Debrecen, Hungary}
\address[2]{Cyclotron Laboratory, Vrije Universiteit Brussel (VUB), Brussels, Belgium}
\address[3]{Institute of Physics and Power Engineering (IPPE), Obninsk, Russia}

\begin{abstract}
Activation cross-sections data of longer-lived products of proton induced nuclear reactions on dysprosium were extended up to 65 MeV by using stacked foil irradiation and gamma spectrometry experimental methods. Experimental cross-sections data for the formation of the radionuclides $^{159}$Dy, $^{157}$Dy, $^{155}$Dy, $^{161}$Tb, $^{160}$Tb, $^{156}$Tb, $^{155}$Tb, $^{154m2}$Tb, $^{154m1}$Tb, $^{154g}$Tb, $^{153}$Tb, $^{152}$Tb and $^{151}$Tb are reported in the 36-65 MeV energy range, and compared with an old dataset from 1964. The experimental data were also compared with the results of cross section calculations of the ALICE and EMPIRE nuclear model codes and of the TALYS nuclear reaction model code as listed in the latest on-line libraries TENDL 2013.
\end{abstract}


\end{frontmatter}


\section{Introduction}
\label{1}
In the frame of this systematic study we have investigated the activation cross sections induced by protons and deuterons on natural dysprosium targets. The part of the study specifically devoted to production of $^{161}$Ho, a candidate therapeutic radioisotope, was published separately \citep{TF2013ARI}. Activation cross-sections of long-lived products of deuteron induced nuclear reactions on dysprosium up to 50 MeV was published in \citep{TF2014a} and of proton induced nuclear reactions on dysprosium up to 36 MeV in \citep{TF2013ANC}. Recently we had the possibility to perform the same measurements at higher energies.  Here we report on the activation cross section data induced by proton irradiation of dysprosium in the 36-65 MeV energy range. Only one set of earlier experimental data were found in the literature \citep{May}.

\section{Experiment and data evaluation}
\label{2}

For the cross section determination an activation method based on stacked foil irradiation technique followed by $\gamma$-ray spectroscopy were used. The general characteristics and procedures for irradiation, activity assessment and data evaluation (including estimation of uncertainties) were similar as in many of our earlier works \citep{TF2013ARI, TF2013ANC,TF2014a}. The stack consisted of a sequence of Al, Dy, Al, Hf, Al, Ti, Be, Ti foils repeated 25 times and bombarded for 3600 s with a 65 MeV proton beam of 25 nA at Louvain la Neuve in one single irradiation. From this irradiation the activation cross sections for production of $^{7}$Be by proton up to 65 MeV were already published, where additional details can be found on experiment on beam parameters and on the data evaluation \citep{Hermanne}. The main experimental parameters and the methods of data evaluation for the present study are summarized in Table 1. The used decay data are collected in Table 2.  

\begin{table*}[t]
\tiny
\caption{Main parameters of the experiment and the methods of data evaluations}
\begin{center}
\begin{tabular}{|p{0.8in}|p{1.6in}|p{1.1in}|p{1.8in}|}
\hline
\multicolumn{2}{|c|}{\textbf{Experiment}}  & \multicolumn{2}{|c|}{\textbf{Data evaluation}} \\
\hline
 & & & \\
\hline
Incident particle & Proton & Gamma spectra evaluation & Genie 2000, 
\citep{Canberra}, Forgamma \citep{Szekely} \\
\hline
Method & Stacked foil & Determination of beam intensity & Faraday cup 
(preliminary)Fitted monitor reaction (final) \citep{TF1991} \\
\hline
Target stack and thicknesses & Al(49.6 $\mu$m), Dy(22.1 $\mu$m), Al(98 $\mu$m),\newline 
Hf(107 $\mu$m), Al(49.6 $\mu$m), Ti(10.9 $\mu$m), \newline Be(285 $\mu$m), Ti(10.9 
v) block \newline Repeated 25 times & Decay data & NUDAT 2.6 \citep{Kinsey} 
\\
\hline
Number of target foils & 25x8 & Reaction Q-values & Q-value calculator 
\citep{Pritychenko}  \\
\hline
Accelerator & Cyclone 90 cyclotron of the Universit\'e Catholique in 
Louvain la Neuve (LLN) Belgium & Determination of beam energy & Andersen 
\citep{Andersen} (preliminary)Fitted monitor reaction (final) 
\citep{TF2001} \\
\hline
Primary energy & 65 MeV & Uncertainty of energy & Cumulative effects of 
possible uncertainties \citep{error} \\
\hline
Irradiation time & 60 min & Cross sections & Elemental cross section \\
\hline
Beam current & 25 nA & Uncertainty of cross sections & Sum in quadrature 
of all individual contribution \\
\hline
Monitor reaction, $[$recommended values$]$ & $^{nat}$Ti(p,x)$^{48
}$V $^{27}$Al(p,x)$^{22,24}$Na reactions & Yield & Physical yield 
\citep{Bonardi} \\
\hline
Monitor target and thickness & $^{nat}$Ti, 10.9 $\mu$m $^{27}$Al 98 $\mu$m 
& Theory & ALICE-IPPE \citep{Dityuk} ,EMPIRE \citep{Herman2007}; 
\citep{KoningTALYS} ,TALYS \citep{KoningTENDL} \\
\hline
detector & HPGe & & \\
\hline
$\gamma$-spectra measurements & 4 series & & \\
\hline
Cooling times (h) & 22.0-31.6 \newline 71.6-96.2\newline 140.6-333.8\newline 2638-3213 & & \\
\hline
\end{tabular}

\end{center}
\end{table*}

\begin{table*}[t]
\tiny
\caption{Decay characteristics of the investigated reaction products and Q-values of reactions for their productions}
\begin{center}
\begin{tabular}{|p{0.4in}|p{0.2in}|p{0.3in}|p{0.2in}|p{0.7in}|p{0.6in}|}
\hline
\textbf{Nuclide} & \textbf{Half-life} & \textbf{E$_{\gamma}$(keV)} & \textbf{I$_{\gamma}$(\%)} & \textbf{Contributing 
reaction} & \textbf{Q-value(keV)} \\
\hline
\textbf{$^{159}$Dy}\newline EC: 100 \% & 144.4 d & 58.0 & 2.27 & $^{160}$
Dy(p,pn)\newline $^{161}$Dy(p,p2n)\newline $^{162}$Dy(p,p3n)\newline $^{163}$Dy(p,p4n)\newline
$^{164}$Dy(p,p5n)\newline $^{159}$Ho decay & 
-8575.9\newline -15030.29\newline -23227.29\newline -29498.3\newline -37156.42\newline -11195.85 \\
\hline
\textbf{$^{157}$Dy}\newline EC:100 \% & 8.14 h & 182.424\newline 326.336 & 1.3393 & 
$^{158}$Dy(p,pn)\newline $^{160}$Dy(p,p3n)\newline $^{161}$Dy(p,p4n)\newline $^{162}$
Dy(p,p5n)\newline $^{163}$Dy(p,p6n)\newline $^{164}$Dy(p,p7n)\newline $^{157}$ Ho decay 
& -9055.54\newline -24464.14\newline -30918.53\newline -39115.52\newline -45386.54\newline -53044.66\newline -12429.3 \\
\hline
\textbf{$^{155}$Dy}\newline EC: 98.62 \% \newline $\beta^{+}$:$^{ }$1.38 \% & 9.9 
h & 184.564\newline 226.918 & 3.37\newline 68.4 & $^{158}$Dy(p,p3n)\newline $^{160}$
Dy(p,p5n)\newline $^{161}$Dy(p,p6n)\newline $^{162}$Dy(p,p7n)\newline $^{163}$Dy(p,p8n)\newline 
$^{164}$Dy(p,p9n)\newline $^{155}$Ho decay & 
-25466.2\newline -40874.7\newline -47329.1\newline -55526.1\newline -61797.1\newline -69455.3\newline -13342.8 \\
\hline
\textbf{$^{161}$Tb}\newline $\beta^{-}$: 100 \% & 6.89 d & 
74.56669\newline 87.941\newline 103.065\newline 106.113\newline 292.401 & 10.2\newline 0.183\newline 0.101\newline 0.078\newline 0.058 & $^{162
}$Dy(p,2p)\newline $^{163}$Dy(p,2pn)\newline $^{164}$Dy(p,2p2n) & 
-8007.59\newline -14278.6\newline -21936.72 \\
\hline
\textbf{$^{160}$Tb}\newline $\beta^{-}$:100 \% & 72.3 d & 
86.7877\newline 298.5783\newline 879.378\newline 966.166\newline 1177.954 & 13.2\newline 26.1\newline 30.1\newline 25.1\newline 14.9 & $^{161
}$Dy(p,2p)\newline $^{162}$Dy(p,2pn)\newline $^{163}$Dy(p,2p2n)\newline $^{164}$
Dy(p,2p3n) & -7507.17\newline -15704.16\newline -21975.17\newline-29633.29 \\
\hline
\textbf{$^{156}$Tb}\newline $\varepsilon$: 100 \% & 5.35 d & 
88.97\newline 199.19\newline 262.54\newline 296.49\newline 356.38\newline 422.34\newline 534.29\newline 1065.11\newline 1154.07\newline 1222.44 & 
18\newline 41\newline 5.8\newline 4.5\newline 13.6\newline 8.0\newline 67\newline 10.8\newline 10.431 & $^{158}$Dy(p,2pn)\newline $^{160}$
Dy(p,2p3n)\newline $^{161}$Dy(p,2p4n)\newline $^{162}$Dy(p,2p5n)\newline $^{163}$
Dy(p,2p6n)\newline $^{164}$Dy(p,2p7n) & 
-15674.87\newline -31083.47\newline -37537.86\newline -45734.85\newline -52005.87\newline -59663.98 \\
\hline
\textbf{$^{155}$Tb}\newline EC: 100 \% & 5.32 d & 
86.55\newline 105.318\newline 148.64\newline 161.29\newline 163.28\newline 180.08\newline 262.27 & 32.0\newline 25.1\newline 2.65\newline 2.76\newline 4.44\newline 7.5\newline 5.3 
& $^{156}$Dy(p,2p)\newline $^{158}$Dy(p,2p2n)\newline $^{160}$Dy(p,2p4n)\newline $^{
161}$Dy(p,2p5n)\newline $^{162}$Dy(p,2p6n)\newline $^{163}$Dy(p,2p7n)\newline $^{164}$
Dy(p,2p8n)\newline $^{155}$Dy decay & 
-6567.84\newline -22589.3\newline -37997.9\newline -44452.3\newline -52649.3\newline -58920.3\newline -66578.4\newline -13342.8 \\
\hline
\textbf{$^{154m2}$Tb}\newline $\varepsilon + \beta^{+}$: 98.2 \%\newline IT: 1.8\%
 & 22.7 h & 123.071\newline 247.925\newline 346.643\newline 992.92\newline 1419.81 & 43\newline 79\newline 69\newline 16.2\newline 46 & $^{156
}$Dy(p,2pn)\newline $^{158}$Dy(p,2p3n)\newline $^{160}$Dy(p,2p5n)\newline $^{161}$
Dy(p,2p6n)\newline $^{162}$Dy(p,2p7n)\newline $^{163}$Dy(p,2p8n) & 
-15732.9\newline -31753.4\newline -47161.3\newline -53615.7\newline -61812.7\newline -68083.7 \\
\hline
\textbf{$^{154m1}$Tb}\newline $\varepsilon + \beta^{+}$:78.2\%\newline IT: 21.8 \% & 9.4 
h & 247.925\newline 540.18\newline 649.564\newline 873.190\newline 996.262 & 22.1\newline 20\newline 10.9\newline 9.2\newline 8.6 & $^{156
}$Dy(p,2pn)\newline $^{158}$Dy(p,2p3n)\newline $^{160}$Dy(p,2p5n)\newline $^{161}$
Dy(p,2p6n)\newline $^{162}$Dy(p,2p7n)\newline $^{163}$Dy(p,2p8n) & 
-15732.9\newline -31753.4\newline -47161.3\newline -53615.7\newline -61812.7\newline -68083.7 \\
\hline
\textbf{$^{154g}$Tb}\newline EC: 97.6\%\newline $\beta^{+}$:2.4 \% & 
21.5 h & 123.07\newline 557.60\newline 722.12\newline 1123.09\newline 1274.436\newline 1291.31 & 26 5.4\newline 7.7\newline 5.7\newline 
10.5\newline 6.9 & $^{156}$Dy(p,2pn)\newline $^{158}$Dy(p,2p3n)\newline $^{160}$
Dy(p,2p5n)\newline $^{161}$Dy(p,2p6n)\newline $^{162}$Dy(p,2p7n)\newline $^{163}$
Dy(p,2p8n) & -15732.9\newline -31753.4\newline -47161.3\newline -53615.7\newline -61812.7\newline -68083.7 \\
\hline
\textbf{$^{153}$Tb}\newline EC: 9.906\% $\beta^{+}$:0.094 \% & 2.34 d & 
102.255\newline 109.75\newline 170.42\newline 212.00 & 6.4\newline 6.8\newline 6.3\newline 31.0 & $^{156}$Dy(p,2p2n)\newline 
$^{158}$Dy(p,2p4n)\newline $^{160}$Dy(p,2p6n)\newline $^{161}$Dy(p,2p7n)\newline$^{162}$Dy(p,2p8n)\newline $^{163}$Dy(p,2p9n)\newline $^{153}$Dy decay & 
-22647.0\newline -38667.46\newline -54075.37\newline -60529.76\newline -68726.74\newline-74997.76\newline-25599.71 \\
\hline
\textbf{$^{152}$Tb}\newline $\varepsilon$: 79.7\%\newline $\beta^{+}$: 20.3 \% & 
17.5 h & 271.09\newline 344.2785\newline 778.9045 & 9.53\newline 63.5\newline 5.54 & $^{156}$Dy(p,2p3n)\newline 
$^{158}$Dy(p,2p5n)\newline $^{160}$Dy(p,2p7n)\newline $^{161}$Dy(p,2p8n)\newline $^{162}$Dy(p,2p9n)\newline $^{152}$Dy decay & 
-31315.0\newline -47335.4\newline -62743.3\newline -69197.7\newline -77394.7\newline -32695.67 \\
\hline
\textbf{$^{151}$Tb}\newline EC: 99.91\%\newline  $\beta^{+}$: 0.99\% \newline $\alpha$: 0.0095\% & 17.609 h & 
108.088\newline 180.186\newline 251.863\newline 287.357\newline 395.444\newline 443.879\newline 479.357\newline 587.46\newline 616.561 & 
24.3\newline 11.5\newline 26.3\newline 28.3\newline 10.8\newline 10.8\newline 15.4\newline 15.6\newline 10.4\newline & $^{156}$Dy(p,2p4n)\newline 
$^{158}$Dy(p,2p6n)\newline $^{160}$Dy(p,2p8n)\newline $^{161}$Dy(p,2p9n)\newline$^{151}$Dy decay & -38479.64\newline-54500.1\newline -69907.99\newline -76362.38\newline -42132.82 \\
\hline
\end{tabular}

\end{center}
\begin{flushleft}
\tiny{\noindent Naturally occurring dysprosium is composed of 7 isotopes ($^{156}$Dy -0.06 \%, $^{158}$Dy-0.10 \%, $^{160}$Dy-2.34 \%, $^{161}$Dy-18.9 \%, $^{162}$Dy-25.5 \%, $^{163}$Dy-24.9 \% and $^{164}$Dy-28.2 \%). \newline
When complex particles are emitted instead of individual protons and neutrons the Q-values have to be decreased by the respective binding energies of the compound particles: np-d, +2.2 MeV; 2np-t, +8.48 MeV; n2p-$^{3}$He, +7.72 MeV; 2n2p-$\alpha$, +28.30 MeV
}
\end{flushleft}
\end{table*}




For beam current and beam energy monitoring, for energy degradation and recoil catcher Al and Ti foils were incorporated downstream of each dysprosium foils in the stack. All monitor foil data were considered simultaneously in order to obtain the beam current and beam energy in each target foil by comparison with the IAEA recommended monitor data \citep{TF2001}. The measured cross sections of the monitor reaction $^{27}$Al(p,x)$^{22,24}$Na and the recommended data are shown in Fig. 1.
\begin{figure}
\includegraphics[width=0.5\textwidth]{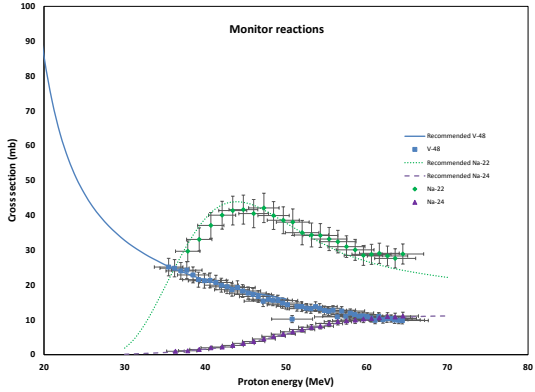}
\caption{The simultaneously measured monitor reactions for determination of proton beam energy and intensity}
\label{fig:1}       
\end{figure}

\section{Results and discussion}
\label{3}
\subsection{Cross sections}
\label{3.1}
The measured cross sections for the production of $^{159}$Dy, $^{157}$Dy, $^{155}$Dy, $^{161}$Tb, $^{160}$Tb, $^{156}$Tb, $^{155}$Tb, $^{154m2}$Tb, $^{154m1}$Tb, $^{154g}$Tb, 153Rb, $^{152}$Tb and $^{151}$Tb   are shown in Table 3-4 and Figures 2-14. The figures also show the theoretical results calculated with the ALICE-IPPE and the EMPIRE codes and the values available in the TALYS based TENDL-2013 library in comparison with experimental results of this work.  In this case we also used the opportunity to compare the results of the older EMPIRE-IPPE version with the newest EMPIRE 3.2 (Malta) \citep{Herman2014} calculations. Due to the experimental circumstances (stacked foil technique, large dose at EOB, limited detector capacity) no cross section data were obtained for short-lived Ho activation products.  Naturally occurring dysprosium is composed of 7 stable isotopes ($^{156}$Dy -0.06 \%, $^{158}$Dy-0.10 \%, $^{160}$Dy-2.34 \%, $^{161}$Dy-18.9 \%, $^{162}$Dy-25.5 \%, $^{163}$Dy-24.9 \% and $^{164}$Dy-28.2 \%). The relevant contributing reactions are collected in Table. 2. 
\newline
Radioisotopes of dysprosium\newline
The investigated radioisotopes of dysprosium are produced directly via (p,pxn) reactions and from the decay of the simultaneously produced holmium parent isotopes.

\subsubsection{The $^{nat}$Dy(p,x)$^{159}$Dy reaction}
\label{3.1.1}
The cumulative cross sections of the $^{159}$Dy (T$_{1/2}$ = 144.4 d) contain, apart from the direct production, the contribution from the decay of $^{159}$Ho (T$_{1/2}$ = 33.05 min) as they were measured after nearly complete decay of the parent isotope. A very old dataset of \citep{May}  show acceptable agreement with our new data up to 60 MeV. The agreement with the results of the 3 codes is also acceptable (Fig. 2), the best approximation is given by the TENDL up to 40 MeV. According to the theory the direct production is negligible, especially below 30 MeV.

\begin{figure}
\includegraphics[width=0.5\textwidth]{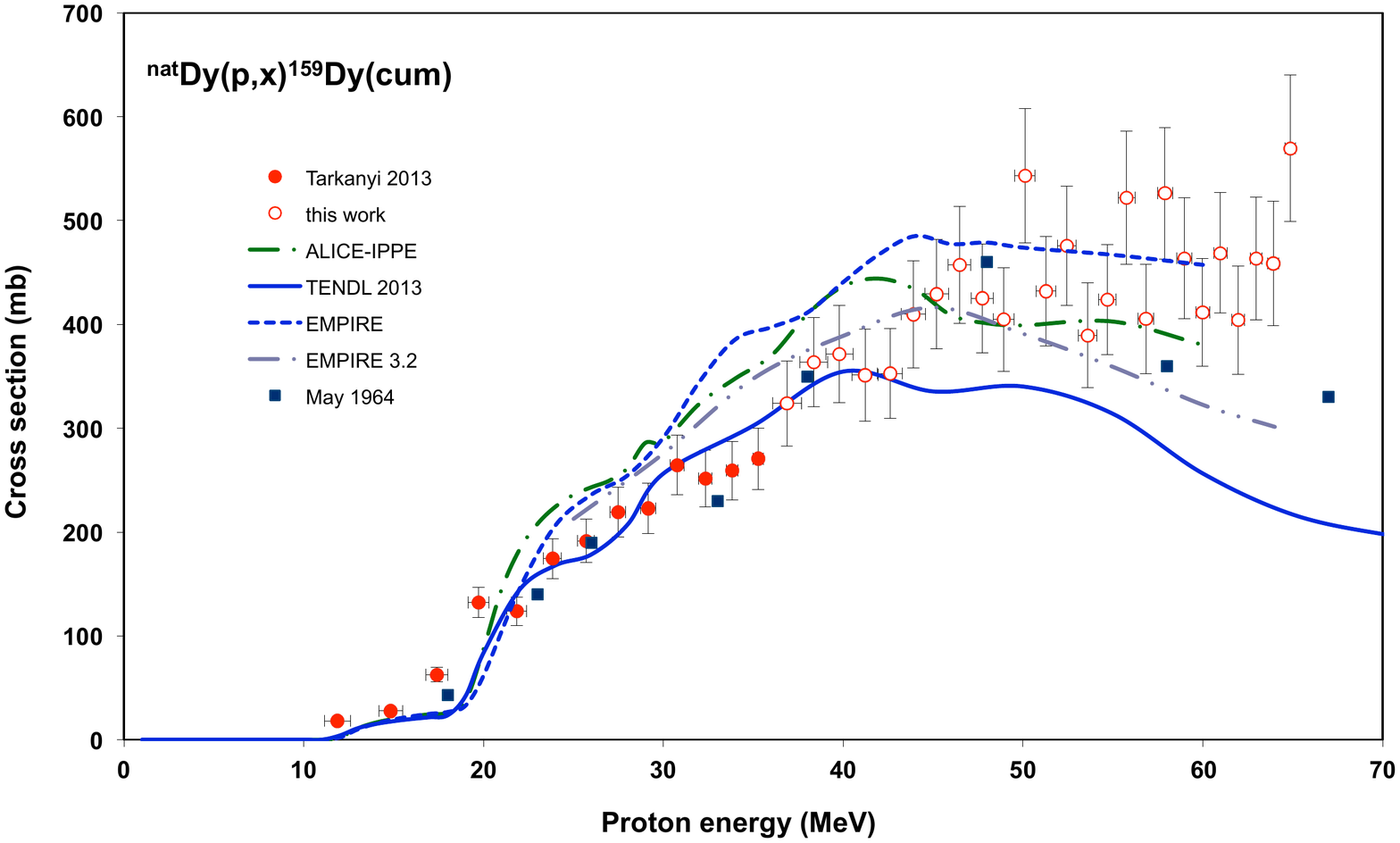}
\caption{Experimental and theoretical cross sections for the formation of $^{159}$Dy by the proton bombardment of dysprosium}
\label{fig:2}       
\end{figure}

\subsubsection{The $^{nat}$Dy(p,x)$^{157}$Dy reaction}
\label{3.1.2}
The cumulative cross sections for production of $^{157}$Dy (T$_{1/2}$ = 8.14 h) was measured after nearly complete decay of the parent $^{157}$Ho (T$_{1/2}$ =12.6 min) (Fig. 3.) The data of May from 1964 are much lower than ours above 40 MeV. The best theoretical approximation is given by the EMPIRE in the whole energy range, ALICE-IPPE also gives good results above 50 MeV, while surprisingly the EMPIRE 3.2 underestimates above 45 MeV. The direct production is negligible. 

\begin{figure}
\includegraphics[width=0.5\textwidth]{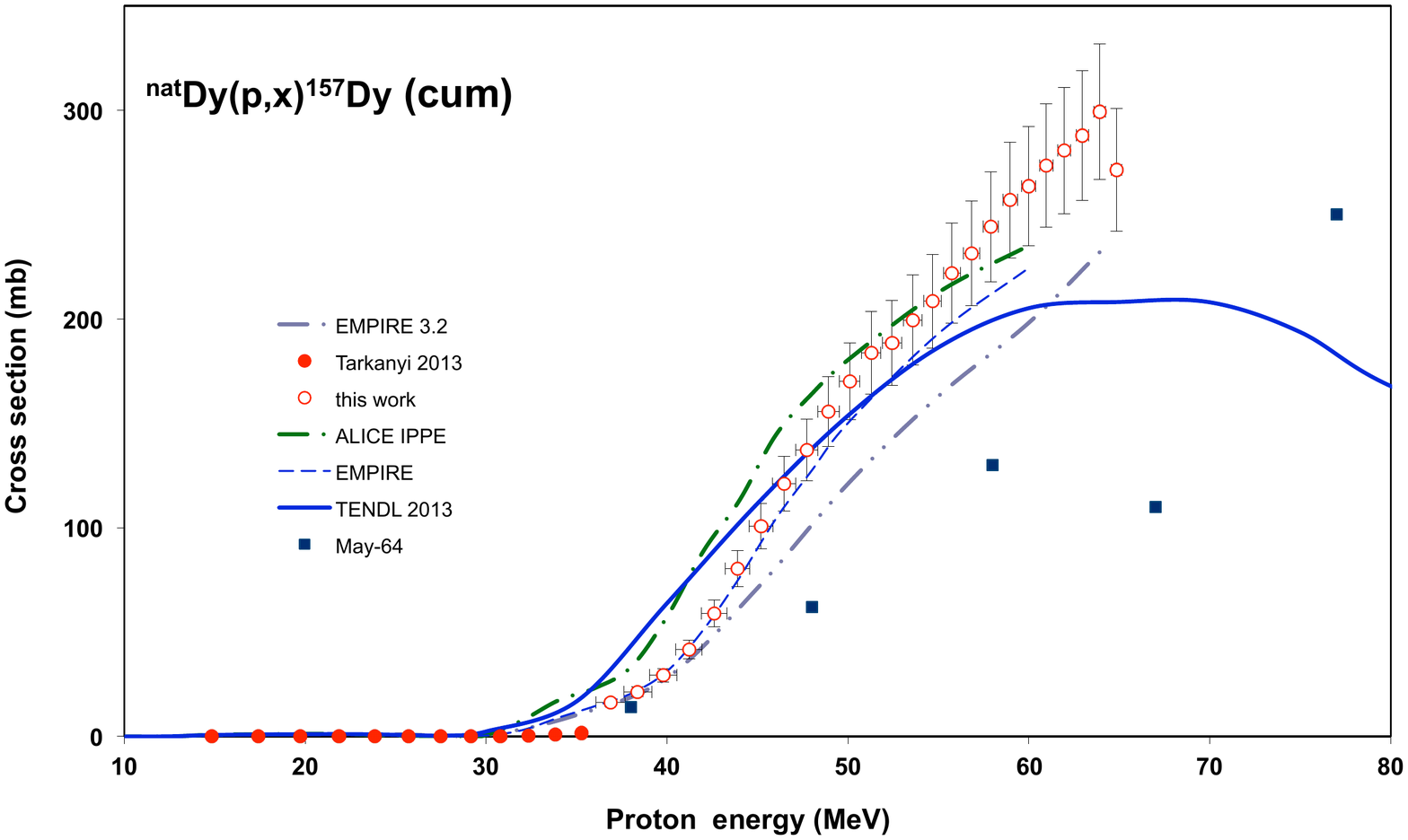}
\caption{Experimental and theoretical cross sections for the formation of $^{157}$Dy by the proton bombardment of dysprosium}
\label{fig:3}       
\end{figure}

\subsubsection{The $^{nat}$Dy(p,x)$^{155}$Dy reaction}
\label{3.1.3}
The measured $^{155}$Dy (T$_{1/2}$ = 9.9 h) was produced directly and through the decay of the $^{155}$Ho, (T$_{1/2}$ = 48 min) parent radioisotope. From the former dataset of May only one point overlaps with our measurements showing an acceptable agreement. The comparison with the theoretical calculations shows disagreement above 50 MeV (Fig. 4), especially TENDL shows a strong overestimation, but the new EMPIRE 3.2 version shows acceptable agreement.

\begin{figure}
\includegraphics[width=0.5\textwidth]{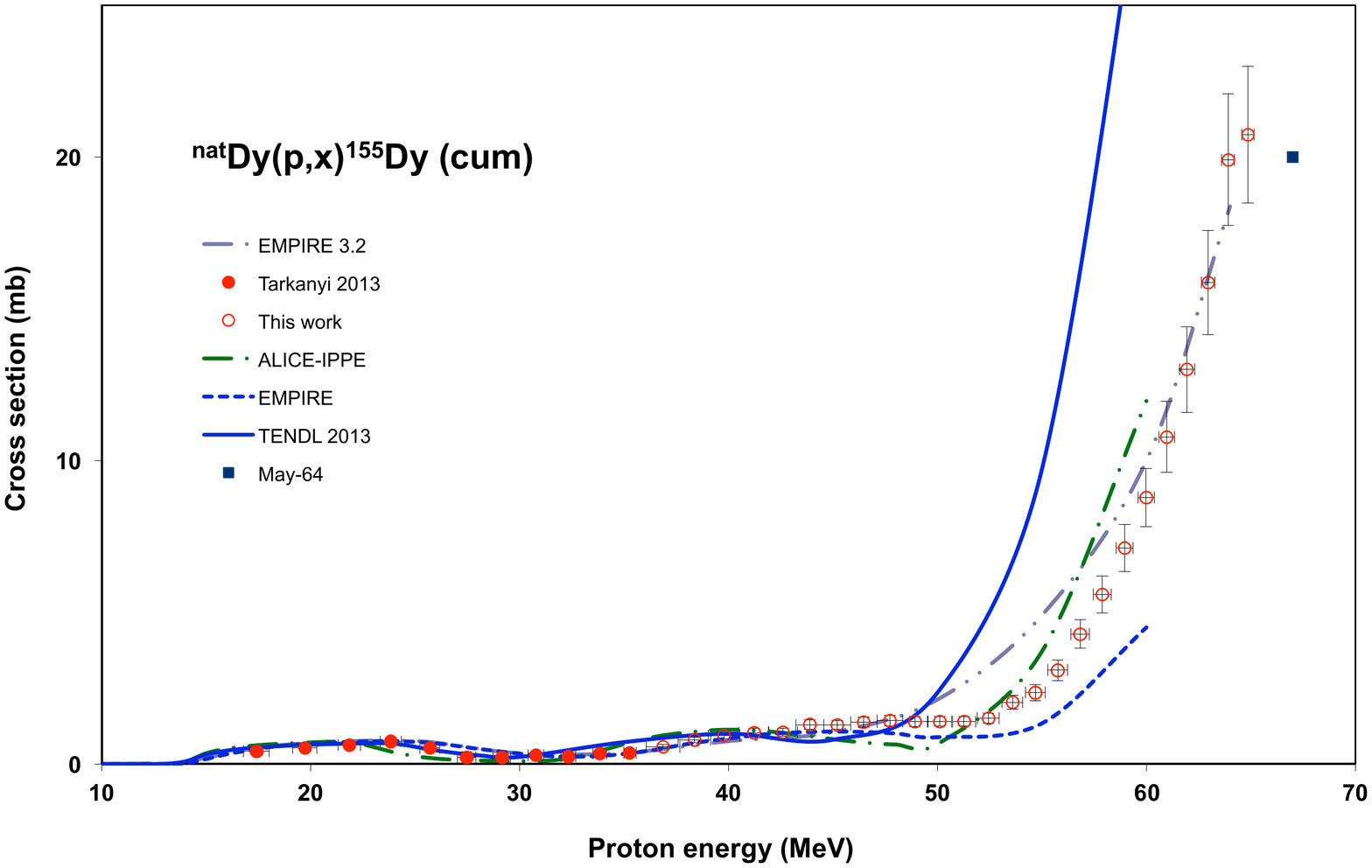}
\caption{Experimental and theoretical cross sections for the formation of $^{155}$Dy by the proton bombardment of dysprosium}
\label{fig:4}       
\end{figure}
\vspace{1 cm}

Radioisotopes of terbium\newline
The terbium radioisotopes are produced through directly (p,2pxn) reactions (including complex particle emissions) and decay of simultaneously produced dysprosium radio-products.

\subsubsection{The $^{nat}$Dy(p,x)$^{161}$Tb reaction}
\label{3.1.4}
The measured direct cross sections for production of $^{161}$Tb (T$_{1/2}$ = 6.89 d) are shown in Fig. 5 in comparison with the theoretical predictions. As can be seen in Table 2, reactions on stable Dy target isotopes with nearly the same abundance can contribute to the production of $^{161}$Tb. From systematics it is known that the (p,2p) reaction has mostly lower cross sections than the (p,2pn) channel. The sharp low energy peak due to $^{162}$Dy(p,2p) that can be seen in the TENDL theoretical data is hence questionable and is reproduced neither by the experimental values, nor by the EMPIRE and ALICE results. In the high-energy range there are large disagreements between the theoretical results. The trend of the experimental data in this energy range support the TENDL calculations, but the new EMPIRE 3.2 gives better approach below 50 MeV. Unfortunately our previous and present data connect at the point, where the cross section begins to increase starkly, as well as because of the low statistic the last points of the present stack are not so reliable, which facts result in a disagreement in the connection range.

\begin{figure}
\includegraphics[width=0.5\textwidth]{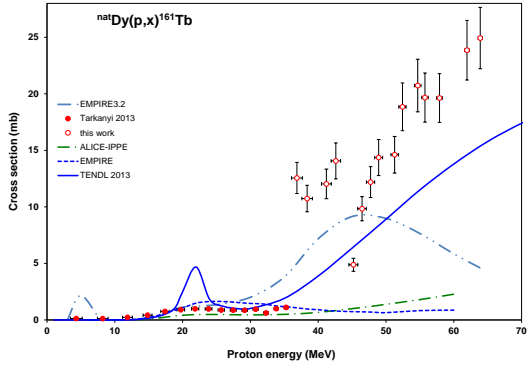}
\caption{Experimental and theoretical cross sections for the formation of $^{161}$Tb by the proton bombardment of dysprosium}
\label{fig:5}       
\end{figure}

\subsubsection{The $^{nat}$Dy(p,xn)$^{160}$Tb reaction}
\label{3.1.5}
The new experimental data of the $^{160}$Tb (T$_{1/2}$ = 72.3 d) are in good agreement with the low energy data in the overlapping energy range. The theoretical descriptions seem to be not very successful (Fig. 6). The sharp peak predicted by the TENDL is not present in the excitation function. The shape of the excitation function is more or less described only by ALICE-IPPE, but the magnitudes are significantly differ from the experimental data in the whole energy range.

\begin{figure}
\includegraphics[width=0.5\textwidth]{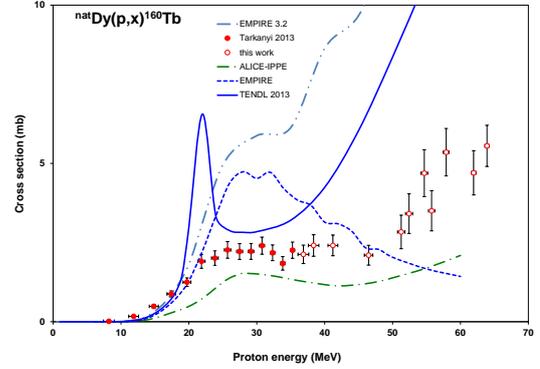}
\caption{Experimental and theoretical cross sections for the formation of $^{160}$Tb by the proton bombardment of dysprosium}
\label{fig:6}       
\end{figure}

\subsubsection{The $^{nat}$Dy(p,xn)$^{156}$Tb reaction}
\label{3.1.6}
The cross sections of $^{156g}$Tb (T$_{1/2}$ = 5.35 d) were obtained from spectra measured after the complete decay of the two short-lived isomeric states (T$_{1/2}$ =5.3 h, IT: 100 \% and T$_{1/2}$ = 24.4 h, IT: 100 \%). In the investigated energy range the ALICE-IPPE represents very closely the experimental data (Fig. 7), the TENDL results underestimate, while both EMPIRE versions overestimate the experimental cross sections by at least a factor of 3.

\begin{figure}
\includegraphics[width=0.5\textwidth]{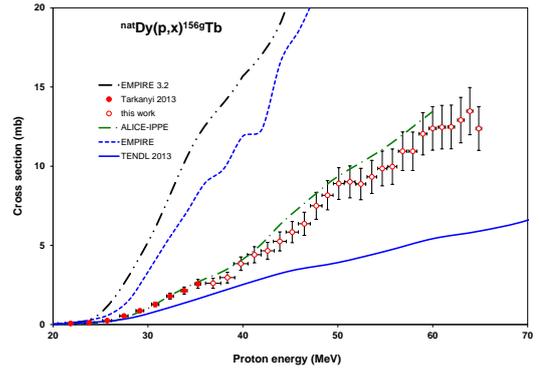}
\caption{Experimental and theoretical cross sections for the formation of $^{156}$Tb by the proton bombardment of dysprosium}
\label{fig:7}       
\end{figure}

\subsubsection{The $^{nat}$Dy(p,xn)$^{155}$Tb reaction}
\label{3.1.7}
The measured cumulative cross sections of $^{155}$Tb (T$_{1/2}$ =5.32 d) contains the complete contribution from the decay of $^{155}$Dy (T$_{1/2}$ =9.9 h). The agreement of by the theories predicted cross sections with the experimental data is shown in Fig. 8.  Only the ALICE-IPPE and TENDL codes give acceptable approximation above 55 MeV.

\begin{figure}
\includegraphics[width=0.5\textwidth]{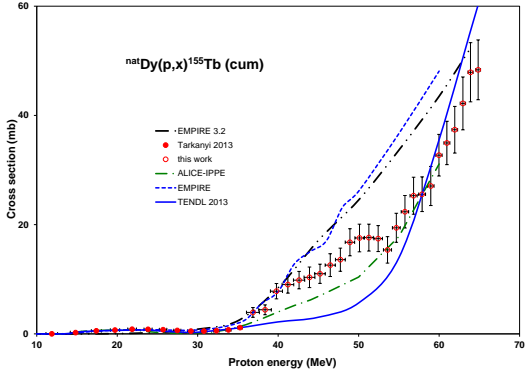}
\caption{Experimental and theoretical cross sections for the formation of 155 Tb by the proton bombardment of dysprosium}
\label{fig:8}       
\end{figure}

\subsubsection{The $^{nat}$Dy(p,xn)$^{154m2}$Tb reaction}
\label{3.1.8}
The $^{154}$Tb has three long-lived isomeric states. The 22.7 h half-life $^{154m2}$Tb isomeric state ($J^{\pi} = 0^+$), decaying with 98.2 \% $\varepsilon + \beta^+$ to $^{154}$Gd, and with 1.8\% IT to $^{154m1}$Tb. The 9.4 h half-life lower laying $^{154m1}$Tb isomeric state ($J^{\pi} = 3^-$) decaying with 78.2 \% $\varepsilon + \beta^+$ to $^{154}$Gd  and with 21.8\% IT to $^{154}$Gdy. The 21.2 h half-life ground state ($J^{\pi} = 7^-$) decays with 100 \% $\varepsilon + \beta^+$ to the stable $^{154}$Gd. For production of isomeric states no contribution from the neighboring isobars: $^{154}$Dy (alpha emitter) and the $^{154}$Gd (stable). Therefore the $^{154m2}$Tb isomeric state is produced only directly via (p,2pxn) reactions. The experimental and theoretical cross sections are shown in Fig. 9. The theoretical model codes overestimate the experimental data. In the case of the theoretical calculations of the two isomer states the main problem is connected to the too complex scheme of low-lying levels of $^{154}$Tb. There are big gaps at the scheme of gamma-transitions between the lowest levels and it effects directly on the calculated results.

\begin{figure}
\includegraphics[width=0.5\textwidth]{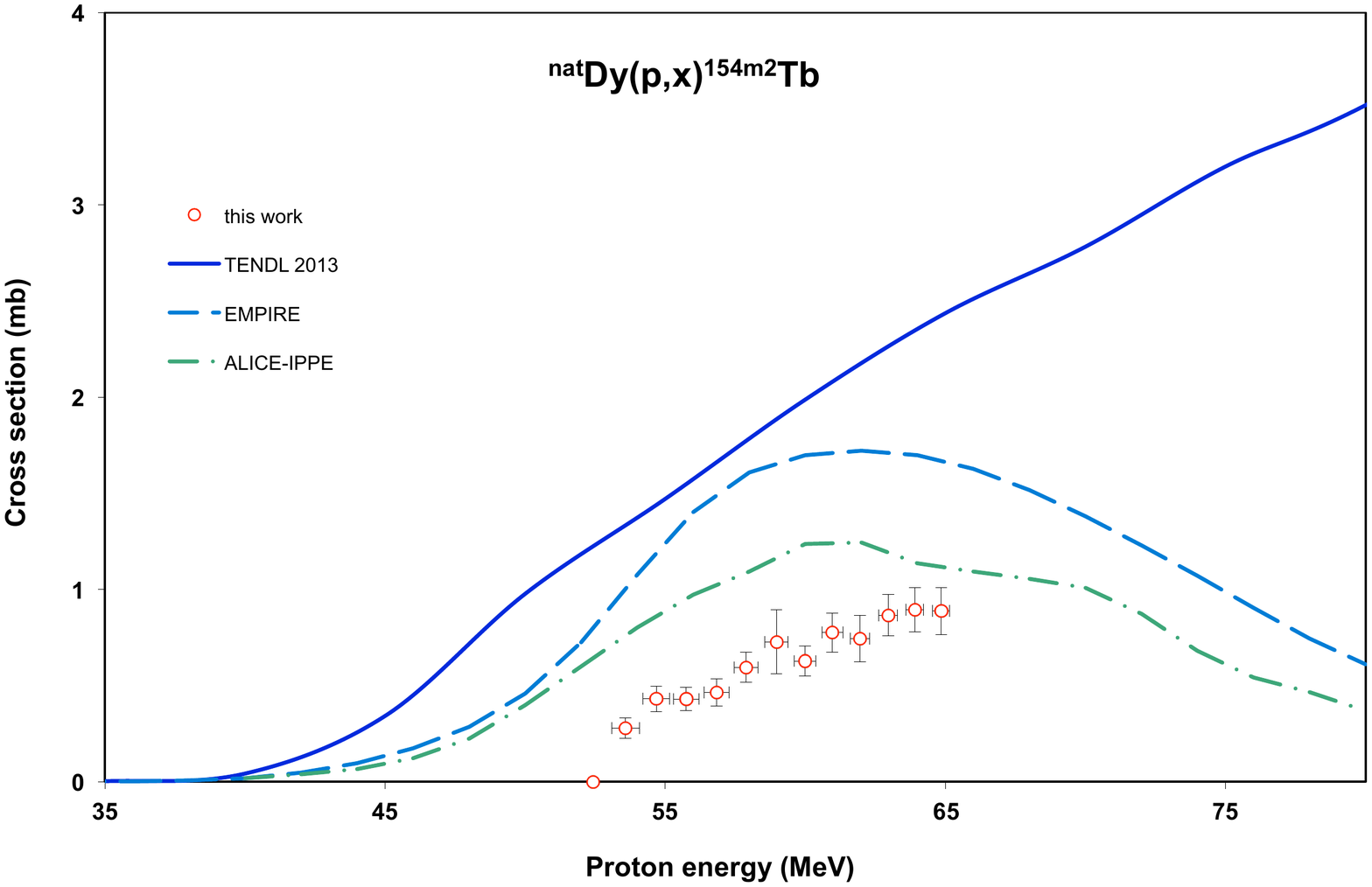}
\caption{Experimental and theoretical cross sections for the formation of $^{154m2}$Tb by the proton bombardment of dysprosium}
\label{fig:9}       
\end{figure}

\subsubsection{The $^{nat}$Dy(p,xn)$^{154m1}$Tb reaction}
\label{3.1.9}
The $^{154m1}$Tb (T$_{1/2}$ = 9.2 h) is produced directly and via a small branching (1.8 \%) from the decay of the $^{154m2}$Tb (22.4 h). Taking into account that our spectra were measured after significant cooling time, the effect of the contribution from the isomeric transition can be significant. Therefore the contribution was subtracted. The direct production cross sections are shown in Fig. 10. If we suppose that up to the first spectra around half of the $^{154m2}$Tb were decayed, but having only 1.8 \%, IT = 0.018*0.5 means only 0.9 \% of the $^{154m2}$Tb. By supposing 9.2 h half-life for this contribution it gives around 3*0.9 = 2.7\% of the $^{154m2}$Tb, i.e. 0.03 mb contribution. The theoretical codes underestimate the experimental values and even do not give the same trend. In the cases of $^{154}$Tb unfortunately EMPIRE 3.2 does not give isomeric information.

\begin{figure}
\includegraphics[width=0.5\textwidth]{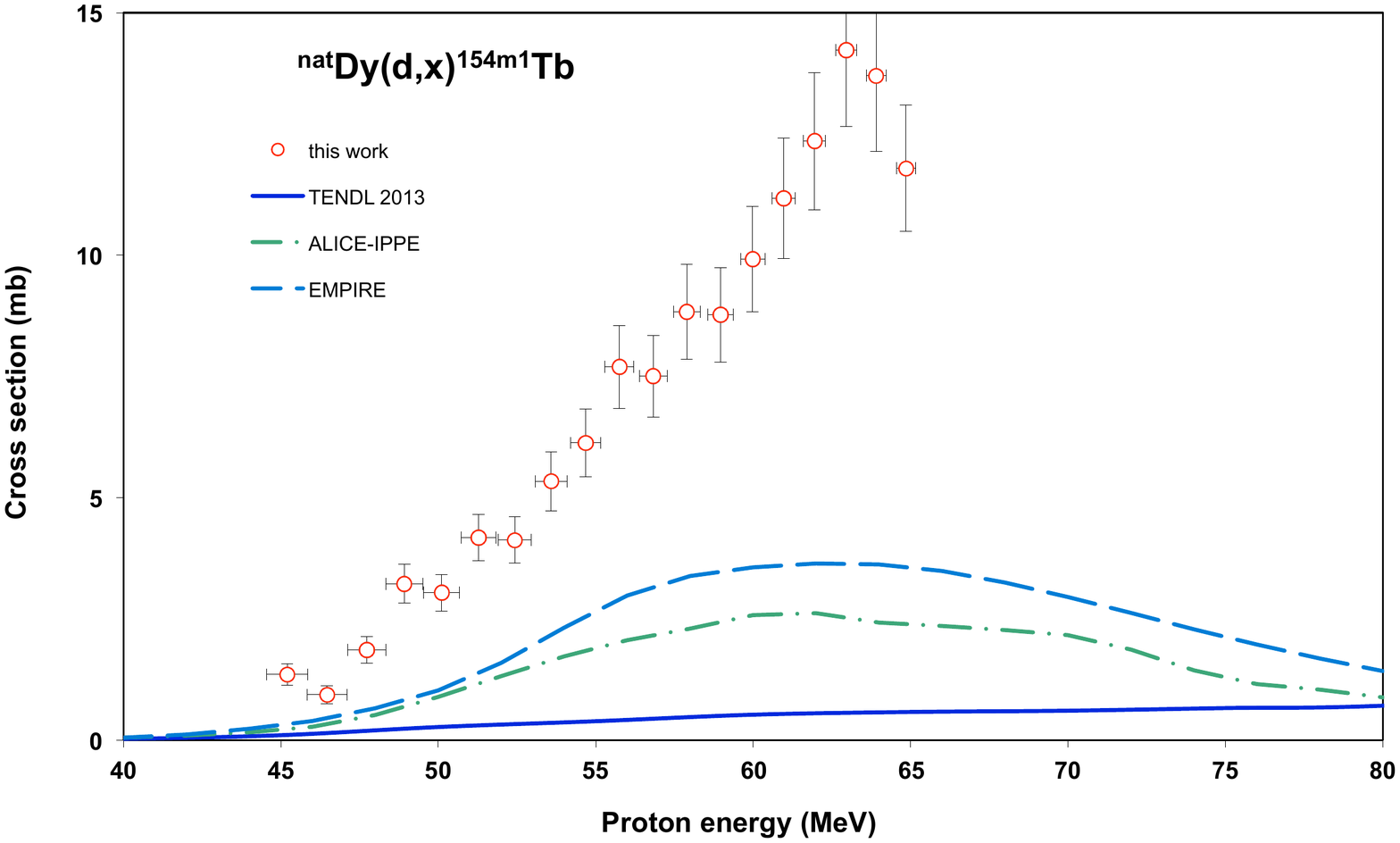}
\caption{Experimental and theoretical cross sections for the formation of $^{154m1}$Tb by the proton bombardment of dysprosium}
\label{fig:10}       
\end{figure}

\subsubsection{The $^{nat}$Dy(p,xn)$^{154g}$Tb reaction (cum)}
\label{3.1.10}
The cross sections were deduced from the first series of spectra measured for around half an hour started in the 22.0-31.6 h after EOB. The results were corrected with the values for $^{154m1}$Tb, because it was not completely disintegrated into the ground state in the time of the measurements. The experimental results and the nuclear reaction model calculations are given in Fig. 11. The TENDL-2013 approximation gives a bit lower values, all other codes strongly overestimate.

\begin{figure}
\includegraphics[width=0.5\textwidth]{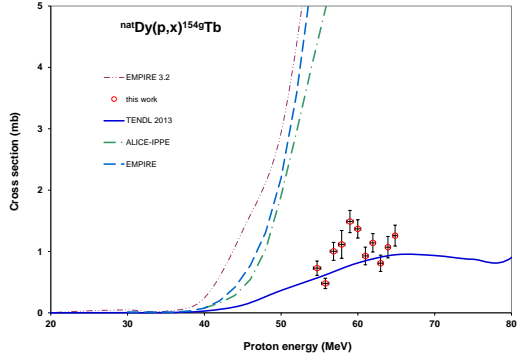}
\caption{Experimental and theoretical cross sections for the formation of $^{154g}$Tb by the proton bombardment of dysprosium}
\label{fig:11}       
\end{figure}

\subsubsection{The $^{nat}$Dy(p,xn)$^{153}$Tb reaction}
\label{3.1.11}
The cumulative cross sections of the $^{153}$Tb (T$_{1/2}$ = 2.34 d) were obtained from spectra measured after complete decay of the $^{153}$Dy parent (T$_{1/2}$ = 6.4 h). The agreement with the TALYS(TENDL-2013) data is acceptable, except the high energy region where the theory underestimates the experiment. (Fig. 12). In this case the EMPIRE 3.2 gives the best and acceptable approximation, while the former EMPIRE version and ALICE-IPPE give overestimating values.

\begin{figure}
\includegraphics[width=0.5\textwidth]{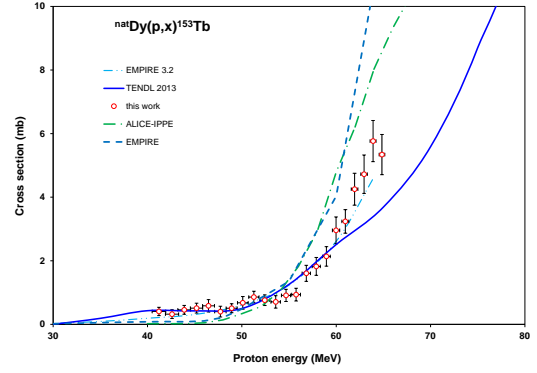}
\caption{Experimental and theoretical cross sections for the formation of $^{153}$Tb  by the proton bombardment of dysprosium}
\label{fig:12}       
\end{figure}

\subsubsection{The $^{nat}$Dy(p,xn)$^{152}$Tb reaction}
\label{3.1.12}
The measured cross sections of the activation cross section of the $^{152}$Tb (T$_{1/2}$ = 17.5 h) ground state additional to direct production include the production through the decay of the $^{152m}$Tb shorter-lived isomeric state (T$_{1/2}$ = 4.2 min, $\varepsilon$: 21.1 \%, IT: 78.9 \%) and the decay of the parent $^{152}$Dy isotope (T$_{1/2}$ = 2.38 h, $\varepsilon$: 99.9 \%, $\alpha$: 0.1 \%). The agreement with theoretical data in the TENDL library (Fig. 13) is good, up to 60 MeV, from which energy ALICE-IPPE gives better results.

\begin{figure}
\includegraphics[width=0.5\textwidth]{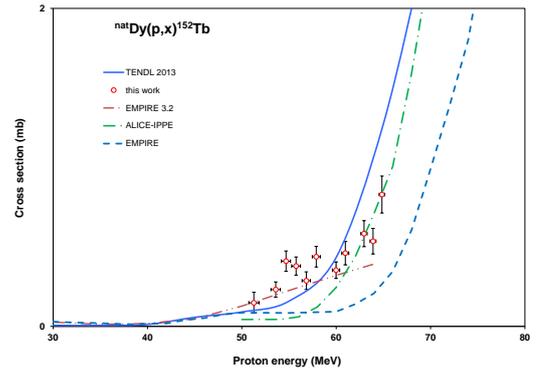}
\caption{Experimental and theoretical cross sections for the formation of $^{152}$Tb by the proton bombardment of dysprosium}
\label{fig:13}       
\end{figure}

\subsubsection{The $^{nat}$Dy(p,xn)$^{151}$Tb reaction}
\label{3.1.13}
We obtained only a few cross section data for cumulative cross section of $^{151}$Tb (T$_{1/2}$ = 17.609 h) near the effective threshold (Fig. 14). The cross sections include the internal decay of the short-lived $^{151m}$Tb (T$_{1/2}$ = 25 s, IT: 93.4 \%, $\varepsilon$: 6.6 \%), and of the parent $^{151}$Dy (T$_{1/2}$ =17.9 min, $\alpha$: 5.6 \%, $\varepsilon$: 94.4 \%). From the theoretical model calculations only the TALYS(TENDL-2013) results show acceptable agreement with the experimental results.

\begin{figure}
\includegraphics[width=0.5\textwidth]{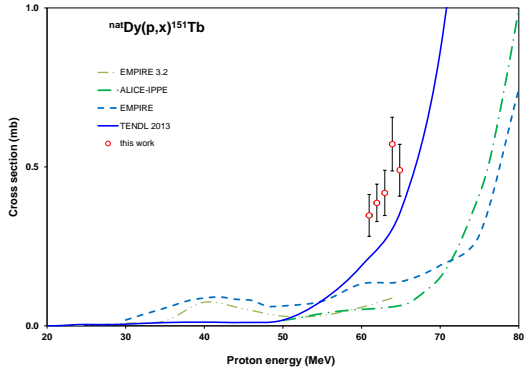}
\caption{Experimental and theoretical cross sections for the formation of $^{151}$Tb by the proton bombardment of dysprosium}
\label{fig:14}       
\end{figure}

\begin{table*}[t]
\tiny
\caption{Experimental cross sections of $^{nat}$Dy(p,xn)$^{159}$Dy, $^{157}$Dy, $^{155}$Dy, $^{161}$Tb, $^{160}$Tb, $^{156}$Tb, and $^{155}$Tb reactions}
\begin{center}
\begin{tabular}{|l|l|l|l|l|l|l|l|l|l|l|l|l|l|l|l|}
\hline
\multicolumn{2}{|c|}{\textbf{$E\pm \delta E$ (MeV)}} & \multicolumn{2}{|c|}{\textbf{$^{159}$Dy}} & \multicolumn{2}{|c|}{\textbf{$^{157}$
Dy}} & \multicolumn{2}{|c|}{\textbf{$^{155}$Dy}} & \multicolumn{2}{|c|}{\textbf{$^{161}$Tb}} & \multicolumn{2}{|c|}{\textbf{
$^{160}$Tb}} & \multicolumn{2}{|c|}{\textbf{$^{156}$Tb}} & \multicolumn{2}{|c|}{\textbf{$^{155}$Tb}} \\
\hline
\multicolumn{2}{|c|}{ } & \multicolumn{14}{|c|}{\textbf{$\sigma \pm \delta\sigma(mb)$}} \\
\hline
36.9 & 0.8 & 323.9 & 41.0 & 16.3 & 1.8 & 0.6 & 0.1 & 12.6 & 1.4 & 2.1 & 
0.3 & 2.6 & 0.3 & 3.9 & 0.9 \\
\hline
38.4 & 0.8 & 363.5 & 43.1 & 21.4 & 2.3 & 0.8 & 0.1 & 10.7 & 1.2 & 2.4 & 
0.3 & 3.0 & 0.3 & 4.4 & 0.9 \\
\hline
39.8 & 0.8 & 371.7 & 46.9 & 29.4 & 3.2 & 0.9 & 0.1 &  & &  & & 3.8 & 
0.4 & 7.8 & 1.4 \\
\hline
41.2 & 0.7 & 351.1 & 44.3 & 41.7 & 4.5 & 1.0 & 0.1 & 12.0 & 1.3 & 2.4 & 
0.3 & 4.4 & 0.5 & 9.0 & 1.6 \\
\hline
42.6 & 0.7 & 352.7 & 43.3 & 58.9 & 6.4 & 1.0 & 0.1 & 14.1 & 1.6 &  & & 
4.7 & 0.5 & 9.8 & 1.6 \\
\hline
43.9 & 0.7 & 409.7 & 51.4 & 80.4 & 8.7 & 1.3 & 0.2 &  & &  & & 5.2 & 
0.6 & 10.4 & 1.9 \\
\hline
45.2 & 0.7 & 429.3 & 52.5 & 100.7 & 10.9 & 1.3 & 0.2 & 4.9 & 0.6 &  & & 
5.8 & 0.7 & 11.0 & 1.8 \\
\hline
46.5 & 0.6 & 457.3 & 56.4 & 121.1 & 13.1 & 1.4 & 0.2 & 9.8 & 1.1 & 2.1 & 
0.3 & 6.4 & 0.7 & 12.6 & 2.1 \\
\hline
47.7 & 0.6 & 425.2 & 52.4 & 137.3 & 14.9 & 1.4 & 0.2 & 12.2 & 1.4 &  & 
& 7.5 & 0.9 & 13.6 & 2.1 \\
\hline
48.9 & 0.6 & 404.8 & 50.0 & 155.6 & 16.8 & 1.4 & 0.2 & 14.4 & 1.6 &  & 
& 8.2 & 0.9 & 16.8 & 2.5 \\
\hline
50.1 & 0.6 & 543.3 & 64.9 & 170.1 & 18.4 & 1.4 & 0.2 &  & &  & & 8.9 & 
1.0 & 17.6 & 2.5 \\
\hline
51.3 & 0.5 & 432.0 & 52.7 & 183.9 & 19.9 & 1.4 & 0.2 & 14.6 & 1.6 & 2.8 
& 0.5 & 9.0 & 1.0 & 17.6 & 2.5 \\
\hline
52.4 & 0.5 & 475.7 & 57.6 & 188.5 & 20.4 & 1.5 & 0.2 & 18.9 & 2.1 & 3.4 
& 0.6 & 8.9 & 1.0 & 17.4 & 2.4 \\
\hline
53.6 & 0.5 & 389.5 & 50.4 & 199.5 & 21.6 & 2.0 & 0.2 &  & &  & & 9.3 & 
1.0 & 15.4 & 2.4 \\
\hline
54.7 & 0.5 & 423.9 & 53.2 & 208.5 & 22.6 & 2.4 & 0.3 & 20.7 & 2.3 & 4.7 
& 0.7 & 9.9 & 1.1 & 19.4 & 2.7 \\
\hline
55.8 & 0.5 & 522.0 & 64.3 & 221.9 & 24.0 & 3.1 & 0.3 & 19.7 & 2.2 & 3.5 
& 0.6 & 10.0 & 1.1 & 22.4 & 3.0 \\
\hline
56.8 & 0.4 & 405.2 & 52.8 & 231.4 & 25.0 & 4.3 & 0.5 &  & &  & & 10.9 
& 1.2 & 25.3 & 3.4 \\
\hline
57.9 & 0.4 & 526.4 & 63.3 & 244.2 & 26.4 & 5.6 & 0.6 & 19.6 & 2.2 & 5.4 
& 0.7 & 10.9 & 1.2 & 25.6 & 3.2 \\
\hline
59.0 & 0.4 & 463.7 & 58.4 & 257.0 & 27.8 & 7.1 & 0.8 &  & &  & & 12.0 
& 1.3 & 27.1 & 3.6 \\
\hline
60.0 & 0.4 & 411.6 & 52.1 & 263.6 & 28.5 & 8.8 & 1.0 &  & &  & & 12.4 
& 1.4 & 32.7 & 3.8 \\
\hline
61.0 & 0.4 & 468.8 & 58.0 & 273.6 & 29.6 & 10.8 & 1.2 &  & &  & & 12.5 
& 1.4 & 35.0 & 4.0 \\
\hline
61.9 & 0.4 & 404.2 & 52.3 & 280.7 & 30.4 & 13.0 & 1.4 & 23.9 & 2.6 & 4.7 
& 0.7 & 12.5 & 1.4 & 37.4 & 4.3 \\
\hline
63.0 & 0.3 & 463.6 & 59.0 & 287.9 & 31.1 & 15.9 & 1.7 &  & &  & & 12.9 
& 1.4 & 42.2 & 4.8 \\
\hline
63.9 & 0.3 & 458.8 & 60.1 & 299.3 & 32.4 & 19.9 & 2.2 & 24.9 & 2.7 & 5.6 
& 0.7 & 13.5 & 1.5 & 47.9 & 5.4 \\
\hline
64.9 & 0.3 & 569.7 & 70.8 & 271.4 & 29.4 & 20.7 & 2.3 & 5.4 & 0.7 &  & 
& 12.4 & 1.4 & 48.3 & 5.5 \\
\hline
\end{tabular}

\end{center}
\end{table*}

\begin{table*}[t]
\tiny
\caption{Experimental cross sections $^{154m2}$Tb, $^{154m1}$Tb, $^{154g}$Tb, $^{153}$Tb, $^{152}$Tb and $^{151}$Tb reactions}
\begin{center}
\begin{tabular}{|l|l|l|l|l|l|l|l|l|l|l|l|l|l|}
\hline
\multicolumn{2}{|c|}{\textbf{$E\pm \delta E$ (MeV)}} & \multicolumn{2}{|c|}{\textbf{$^{154m2}$Tb}} & \multicolumn{2}{|c|}{\textbf{$^{154m1
}$Tb}} & \multicolumn{2}{|c|}{\textbf{$^{154g}$Tb}} & \multicolumn{2}{|c|}{\textbf{$^{153}$Tb}} & 
\multicolumn{2}{|c|}{\textbf{$^{152}$Tb}} & \multicolumn{2}{|c|}{\textbf{$^{151}$Tb}} \\
\hline
\multicolumn{2}{|c|}{ } & \multicolumn{12}{|c|}{\textbf{$\sigma \pm \delta\sigma(mb)$}} \\
\hline
36.9 & 0.8 &  & &  & &  & &  & &  & &  & \\
\hline
38.4 & 0.8 &  & &  & &  & &  & &  & &  & \\
\hline
39.8 & 0.8 &  & &  & &  & & 1.07 & 0.28 &  & &  & \\
\hline
41.2 & 0.7 &  & &  & &  & & 0.41 & 0.13 &  & &  & \\
\hline
42.6 & 0.7 &  & &  & &  & & 0.32 & 0.14 &  & &  & \\
\hline
43.9 & 0.7 &  & &  & &  & & 0.46 & 0.14 &  & &  & \\
\hline
45.2 & 0.7 &  & & 1.35 & 0.22 &  & & 0.51 & 0.15 &  & &  & \\
\hline
46.5 & 0.6 &  & & 0.94 & 0.19 &  & & 0.58 & 0.19 &  & &  & \\
\hline
47.7 & 0.6 &  & & 1.86 & 0.27 &  & & 0.40 & 0.17 &  & &  & \\
\hline
48.9 & 0.6 &  & & 3.22 & 0.40 &  & & 0.51 & 0.14 &  & &  & \\
\hline
50.1 & 0.6 &  & & 3.04 & 0.38 &  & & 0.68 & 0.20 &  & &  & \\
\hline
51.3 & 0.5 &  & & 4.18 & 0.48 &  & & 0.86 & 0.18 & 0.15 & 0.07 &  & 
\\
\hline
52.4 & 0.5 &  & & 4.13 & 0.48 &  & & 0.76 & 0.17 &  & &  & \\
\hline
53.6 & 0.5 & 0.28 & 0.05 & 5.34 & 0.61 &  & & 0.71 & 0.20 & 0.23 & 0.05 
&  & \\
\hline
54.7 & 0.5 & 0.43 & 0.07 & 6.13 & 0.70 & 0.73 & 0.14 & 0.91 & 0.19 & 
0.41 & 0.06 &  & \\
\hline
55.8 & 0.5 & 0.43 & 0.06 & 7.70 & 0.86 & 0.48 & 0.10 & 0.94 & 0.20 & 
0.38 & 0.06 &  & \\
\hline
56.8 & 0.4 & 0.46 & 0.07 & 7.51 & 0.84 & 1.00 & 0.18 & 1.60 & 0.25 & 
0.29 & 0.06 &  & \\
\hline
57.9 & 0.4 & 0.59 & 0.08 & 8.83 & 0.98 & 1.12 & 0.27 & 1.82 & 0.28 & 
0.44 & 0.06 &  & \\
\hline
59.0 & 0.4 & 0.73 & 0.17 & 8.77 & 0.98 & 1.49 & 0.22 & 2.14 & 0.31 &  & 
&  & \\
\hline
60.0 & 0.4 & 0.63 & 0.08 & 9.92 & 1.08 & 1.37 & 0.18 & 2.96 & 0.42 & 
0.35 & 0.05 &  & \\
\hline
61.0 & 0.4 & 0.78 & 0.10 & 11.2 & 1.2 & 0.93 & 0.17 & 3.24 & 0.37 & 0.46 
& 0.07 & 0.35 & 0.07 \\
\hline
61.9 & 0.4 & 0.74 & 0.12 & 12.4 & 1.4 & 1.14 & 0.18 & 4.25 & 0.50 &  & 
& 0.39 & 0.06 \\
\hline
63.0 & 0.3 & 0.87 & 0.11 & 14.2 & 1.6 & 0.81 & 0.16 & 4.72 & 0.60 & 0.58 
& 0.08 & 0.42 & 0.07 \\
\hline
63.9 & 0.3 & 0.89 & 0.11 & 13.7 & 1.6 & 1.07 & 0.21 & 5.76 & 0.65 & 0.54 
& 0.08 & 0.57 & 0.08 \\
\hline
64.9 & 0.3 & 0.89 & 0.12 & 11.8 & 1.3 & 1.26 & 0.21 & 5.34 & 0.63 & 0.83 
& 0.12 & 0.49 & 0.08 \\
\hline
\end{tabular}

\end{center}
\end{table*}

\subsection{Integral yields}
\label{3.2}
The integral yields calculated from spline fits to our experimental excitation functions are shown in Fig 15 (Dy radioisotopes) and 16. (Tb radioisotopes). The integral yields represent so called physical yields i.e. activity instantaneous production rates \citep{Bonardi}. No experimental thick target yield data were found in the literature for the comparison.

\begin{figure}
\includegraphics[width=0.5\textwidth]{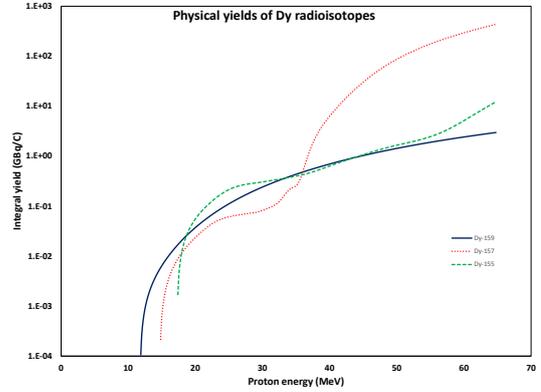}
\caption{Integral thick target yields for the formation of $^{159}$Dy, $^{157}$Dy, $^{155}$Dy in proton induced nuclear reaction on $^{nat}$Dy as a function of the energy}
\label{fig:15}       
\end{figure}

\begin{figure}
\includegraphics[width=0.5\textwidth]{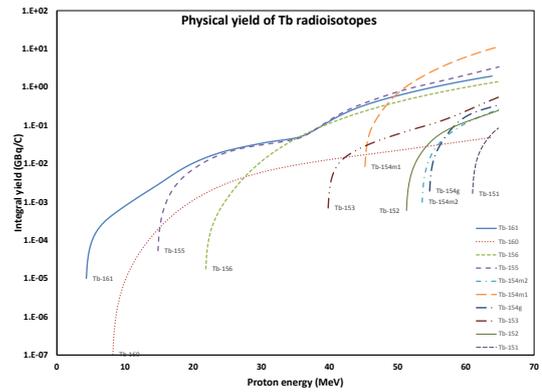}
\caption{Integral thick target yields for the formation of $^{161}$Tb, $^{160}$Tb, $^{156}$Tb, $^{155}$Tb, $^{154m2}$Tb, $^{154m1}$Tb, $^{154g}$Tb, $^{153}$Tb, $^{152}$Tb, $^{151}$Tb in proton induced nuclear reaction on $^{nat}$Dy as a function of the energy}
\label{fig:16}       
\end{figure}

\section{Summary and applications}
\label{4}
We present new experimental cross sections for the $^{nat}$Dy(p,x) $^{159}$Dy, $^{157}$Dy, $^{155}$Dy, $^{161}$Tb, $^{160}$Tb, $^{156}$Tb, $^{155}$Tb, $^{154m2}$Tb, $^{154m1}$Tb, $^{154g}$Tb, $^{153}$Tb, $^{152}$Tb and $^{151}$Tb   nuclear reactions in the  36-65  MeV energy range. The experimental data were compared with the results obtained by the TALYS 1.6 code reported in the TENDL-2013 library and with the results of ALICE-IPPE and two versions of the EMPIRE codes. The theoretical description gives only moderate agreements. The comparisons show that the experimental data are very important for testing the predictive power and improving the performances of the model codes, taking into account that no other experimental activation data are available for these reactions. The experimental data are of importance for several practical applications. Among the investigated reaction products we can mention: 
\begin{itemize} 
\item The radionuclide $^{159}$Dy (T$_{1/2}$ = 144 d, EC =100\%) is a pure Auger electron and X-ray emitter and has gained interest in transmission imaging and bone mineral analysis \citep{Nayak}, while $^{157}$Dy (T$_{1/2}$ =8.14 h.)\citep{Lebowitz}, was investigated as a bone seekers in the evaluation of bone lesions \citep{Hubner}.
\item Terbium offers 4 clinically interesting radioisotopes with complementary physical decay characteristics: $^{149}$Tb, $^{152}$Tb, $^{155}$Tb, and $^{161}$Tb. The identical chemical characteristics of these radioisotopes allows the preparation of radiopharmaceuticals with identical pharmaco-kinetics useful for PET ($^{152}$Tb) or SPECT diagnosis ($^{155}$Tb) and for $\alpha$- ($^{149}$Tb) and $\beta^-$-particle ($^{161}$Tb) therapy \citep{Muller}. From Dy targets the $^{152}$Tb, $^{155}$Tb, and $^{161}$Tb can be obtained, but we showed in a recent publication that other production methods are better \citep{TF2014a,TF2013ANC}.
\item	Radioisotope Gadolinium ($^{153}$Gd) is used to manufacture diagnostic sources of photon radiation, in-line sources and calibration phantoms. Because of its magnetic properties, gadolinium is also applied in intravenous radio-contrast agents in magnetic resonance imaging (MRI). There are broad variety of processes to produce the $^{153}$Gd with high specific activity. The high radioisotope purity (99.97\%) $^{153}$Gd can be obtained at high flux reactors via the  (n,$\gamma$) reaction \citep{Karelin}. The charged particle routes include direct and generator methods. The direct route includes proton and deuteron induced reactions on europium \citep{Nichols,Takacs}  and generator methods via $^{nat}$Gd(p,xn) \citep{Vermeulen} $^{nat}$Gd(d,xn) \citep{TF2014b} and $^{151}$Eu($\alpha$,2n)$^{153}$Tb $\longrightarrow$ $^{153}$Gd \citep{Gedeonov} reactions. In principle the presently investigated $^{nat}$Dy(p,x)$^{153}$Tb $\longrightarrow$ $^{153}$Gd also could be taken into account, but the production cross sections by using $^{nat}$Dy are very low (see Fig. 13) comparing to other indirect charged particle induced reactions.
\end{itemize}

\section{Acknowledgements}
\label{}
This work was performed in the frame of the HAS-FWO Vlaanderen (Hungary-Belgium) project. The authors acknowledge the support of the research project and of the respective institutions. We thank to Cyclotron Laboratory of the Université Catholique in Louvain la Neuve (LLN) providing the beam time and the crew of the LLN Cyclone 90 cyclotron for performing the irradiations. The authors also acknowledge the support from R. Capote (IAEA) in using the newest EMPIRE 3.2 (Malta) version. 
 



\bibliographystyle{elsarticle-harv}
\bibliography{Dyp}







\end{document}